\documentclass[%
 reprint,
 amsmath,amssymb,
 aps,
]{revtex4-1}
\usepackage{amsmath,times,amssymb,latexsym,graphics,braket,color,here}
\usepackage{graphicx}
\usepackage{dcolumn}
\usepackage{bm}
\usepackage{braket}
\usepackage{graphicx} 
\usepackage{color}   

\usepackage{hyperref}

\newcommand{\pa}[1]{\left(#1 \right)}

\newcommand{\ca}[1]{\mathcal{#1}}

\newcommand{\abs}[1]{\left|#1\right|}

\newcommand{\circled}[1]{\textcircled{\scriptsize#1}}

\newcommand{\kett}[1]{\ket{#1\rangle}}

\newcommand{\fr}{\frac}
\newcommand{\s}[1]{\sqrt{#1}}

\def\be{\begin{equation}}
\def\ee{\end{equation}}
\def\ba{\begin{eqnarray}}
\def\ea{\end{eqnarray}}

 \def\ba{{\bar{\alpha}}}

\def\tr{{\text{tr}}}

\makeatletter
\let\cat@comma@active\@empty
\makeatother

\begin{document}
\preprint{RIKEN-iTHEMS-Report-22}
\title{Reflected Entropy in Boundary/Interface Conformal Field Theory}
\date{\today}
\author{Yuya Kusuki}\email[]{ykusuki@caltech.edu}
\affiliation{\it Walter Burke Institute for Theoretical Physics, 
California Institute of Technology, Pasadena, CA 91125, USA.}
\affiliation{\it RIKEN Interdisciplinary Theoretical and Mathematical Sciences (iTHEMS),
Wako, Saitama 351-0198, Japan.}

\begin{abstract}
Boundary conformal field theory (BCFT) and interface conformal field theory (ICFT) attract attention in the context of the information paradox problem.
On this background, we develop the idea of the reflected entropy in BCFT/ICFT.
We first introduce the left-right reflected entropy (LRRE) in BCFT and show that its holographic dual is given by the area of the entanglement wedge cross section (EWCS) through AdS/BCFT.
We also present how to evaluate the reflected entropy in ICFT.
By using this technique, we can show the universal behavior of the reflected entropy in some special classes.
\end{abstract}
\maketitle

\section{Introduction}
The entanglement entropy (EE) plays a significant role in quantum information, condensed matter, and quantum gravity \cite{Ryu2006}.
This quantity captures the bipartite entanglement between a subsystem $A$ and its complement $\bar{A}$.
The EE is defined by the von-Neumann entropy for the reduced density matrix $\rho_A\equiv \tr_{\bar{A}} \rho$ as $S(A)=-\tr \rho_A \log \rho_A$.

One interesting direction to develop this idea is finding a tripartite entanglement measure.
Recently, as one of them, the reflected entropy is introduced by \cite{Dutta2021}.
This quantity is applied to various (1+1)-$d$ setups \cite{Dutta2021,KudlerFlam2021,KudlerFlam2020,Kusuki2021,Kusuki2020,Bueno2020,Zou2022}, (2+1)-$d$ setups \cite{Liu2022, Siva2021, Berthiere2021}, and arbitrary dimensional setups \cite{Bueno2020a,Camargo2021}.
One of the significant features of the reflected entropy
is that this quantity has a nice bulk dual, the minimal area of the cross section in the entanglement wedge \cite{Dutta2021}.
It means that like the EE, one can probe the entanglement structure of quantum gravity by a simple calculation of the area.
Another feature is that
mostly bipartite entanglement patterns imply that the difference between the reflected entropy and the mutual information is close to zero, $S_R-I\simeq 0$ \cite{Akers2020, Zou2021} \footnote{More precisely, $S_R-I\simeq0$ implies no W-like entanglement. It does not imply no GHZ entanglement.}.
Inspired by this sensitivity to tripartite entanglement, the difference $S_R-I$ is studied in \cite{Bueno2020,Bueno2020a,Camargo2021,Hayden2021}, called the Markov gap.

In this article, we develop the idea of the reflected entropy in boundary CFT (BCFT) and interface CFT (ICFT).
BCFT is introduced in \cite{Cardy2004} and developed in many works.
In this article, we particularly focus on the left-right entanglement, which is mainly explored in BCFT \cite{PandoZayas2015, Das2015, Affleck2009}.
Interface CFT (ICFT) is a class of CFTs where two (possibly different) CFTs are connected along an interface \cite{Oshikawa1996,Oshikawa1997,Bachas2002}.
The entanglement entropy in ICFT is studied in various (1+1)-$d$ CFT setups \cite{Sakai2008, Brehm2015,Brehm2016,Wen2018,Gutperle2016a}.
If one considers the reflected entropy in BCFT/ICFT, one may have several questions, for example, can we directly extract the entanglement wedge cross section between the subsystem and the island (see FIG.\ref{fig:EWCS})?
how can we evaluate the reflected entropy in ICFTs?
We answer these questions.
Regarding the first question, we will introduce a new quantity ``left-right reflected entropy (LRRE)'' and show its holographic dual.
In regards to the second question, we will introduce a new technique to evaluate reflected entropy in ICFT/BCFT. This has a wide range of applications. For example, the analysis in \cite{Liu2022} is based on numerical calculation because of a technical reason. We can now overcome this problem by the method developed in this article \cite{KudlerFlam}.

There is another motivation to investigate the reflected entropy in BCFT/ICFT.
Recent progress on the information paradox problem is provided in a class of toy models where the black hole and a non-gravitational bath CFT are glued along the (asymptotic) boundary, which is called the island model \cite{Penington2020,Almheiri2019,Almheiri2020}. This model is related to BCFT/ICFT through the AdS/BCFT and the braneworld holography
\cite{Almheiri2020b,Sully:2020pza,Akal2020,
Bousso:2020kmy,
Miao2021,
Akal2021,
Geng2020,
Geng2021,
Chu:2021gdb,
Akal:2021foz,
Ageev:2021ipd,
Geng2021a,
Kusuki2022,
Suzuki2022,
Numasawa2022,
Izumi2022,
Kusuki2022a}.

There are several works about the reflected entropy in the island model \cite{Akers2022,Chandrasekaran2020,Hayden2021,Li2020,Ghodrati2021}.
In this context, our new measure in BCFT and new technique in ICFT have the potential to provide a new understanding of the island model.

\section{Left-Right Mutual Information}
\begin{figure}[t]
 \begin{center}
  \includegraphics[width=7.0cm,clip]{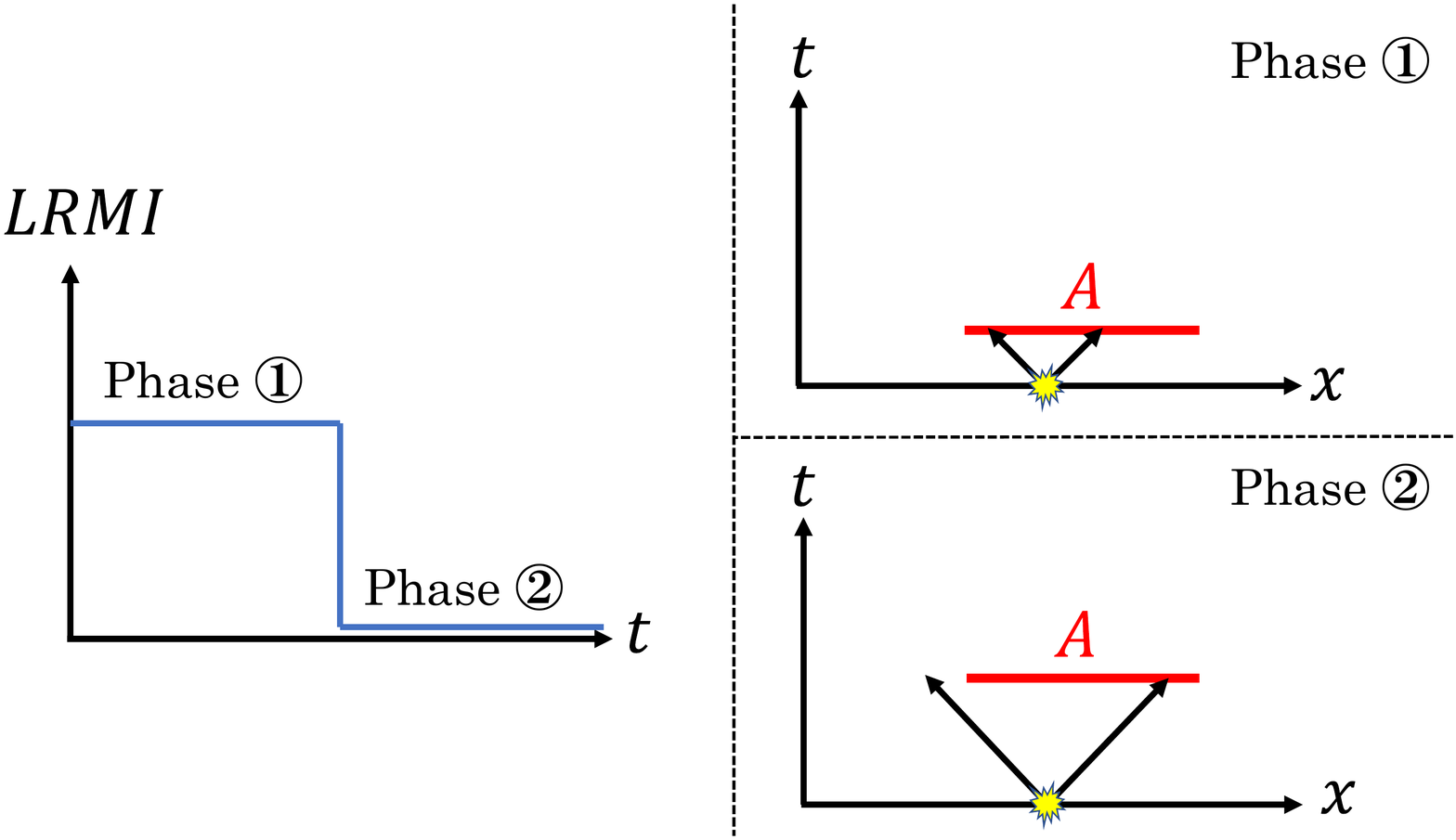}
 \end{center}
 \caption{To give an interpretation of the LRMI, it would be nice to consider the time dependence of the LRMI after a local quench in an integrable system.
 In the phase $\circled{1}$, both the left and right mover are in the subsystem $A$. This entanglement pair increases the LRMI.
 In the phase $\circled{2}$, there is no such a pair in the subsystem $A$. Consequently, the LRMI vanishes.
 }
 \label{fig:local}
\end{figure}

The left-right entanglement entropy (LREE) is defined by a reduced density matrix obtained by tracing over the right moving sector \cite{PandoZayas2015, Das2015, Affleck2009},
\begin{equation}
S^{(l/r)} \equiv -\tr \rho^{(L)} \log  \rho^{(L)},
\end{equation}
where $ \rho^{(L)} \equiv \tr_R \rho$.
One can generalize this quantity by considering a reduced density matrix obtained by tracing over the right moving sector and a part of the left moving sector,
\begin{equation}
S^{(L)}(A) \equiv -\tr \rho^{(L)}_A \log  \rho^{(L)}_A,
\end{equation}
where $\rho^{(L)}_A \equiv \tr_{\bar{A} \cup R} \rho$ .
This quantity can be calculated by the replica trick in the same way as in \cite{PandoZayas2015, Das2015, Affleck2009}.
One can also calculate it by a correlation function with {\it chiral} twist operators \footnote{We can formally define the chiral twist operator by the unwrapping \cite{Lunin2001}. 
For example, the OPE coefficients including the chiral twist operators are evaluated by this prescription.  See also \cite{Hoogeveen2015} }.
With this quantity, one can introduce an interesting quantum information quantity, which we call the {\it left-right mutual information (LRMI)},
\begin{equation}
I^{(l/r)}(A) \equiv S^{(L)}(A) + S^{(R)}(A) - S(A).
\end{equation}
The physical interpretation is in the following.
Let us consider a local excitation that creates a pair of left and right movers.
The LRMI increases if both of them are included in the subsystem $A$.
That is,  the LRMI counts the number of such pairs (see FIG. \ref{fig:local}).

\begin{figure}[t]
 \begin{center}
  \includegraphics[width=7.0cm,clip]{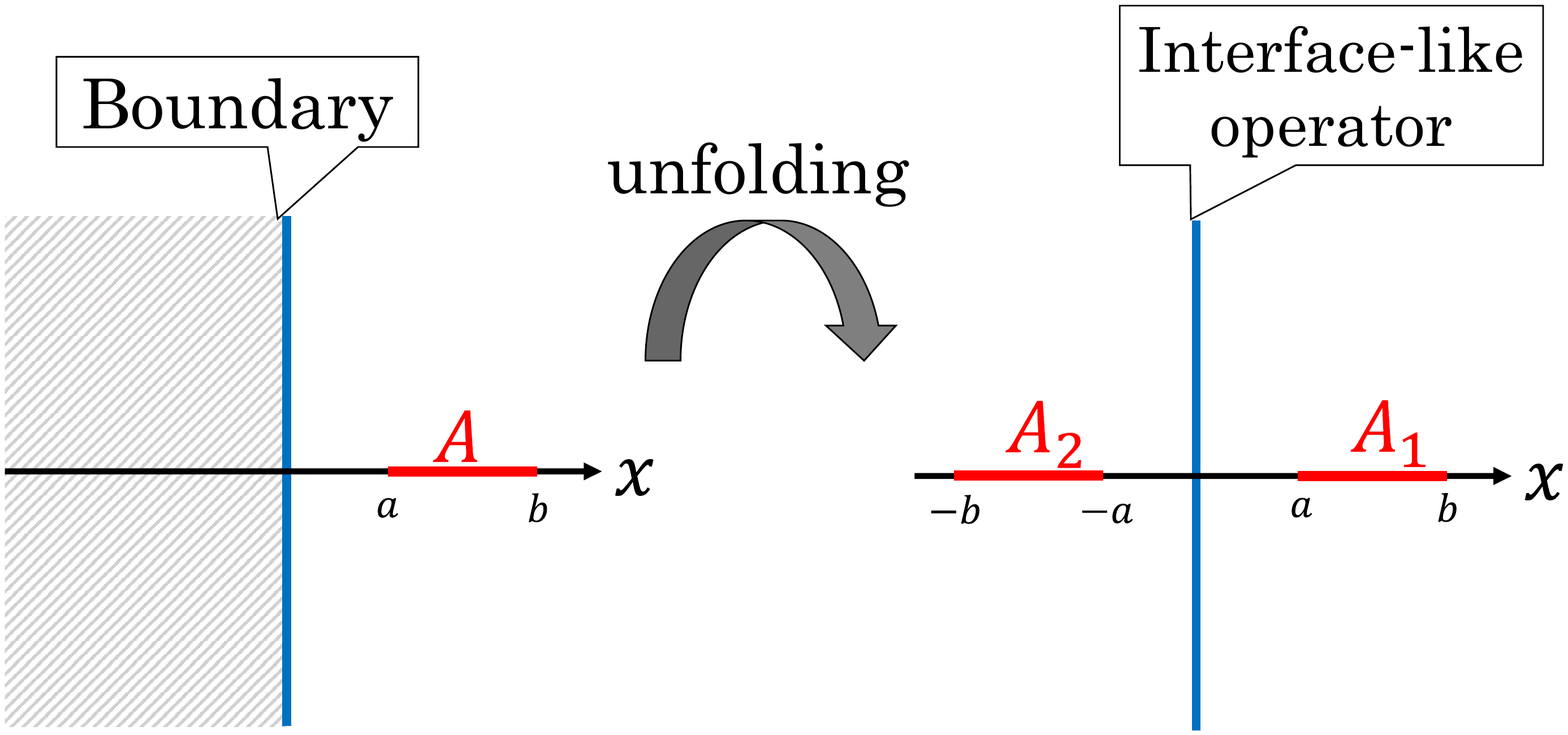}
 \end{center}
 \caption{The unfolding process of the boundary state. The kinematics can be fixed by the conformal symmetry of the full plane, called as the doubling trick \cite{Cardy2004}.}
 \label{fig:unfolding}
\end{figure}

In CFTs with time-like boundary, the LRMI has a nice picture.
Let us consider the LRMI on a half-plane $\Re z > 0$ (see the left of FIG. \ref{fig:unfolding}).
In a similar way to the state-operator correspondence, a conformal boundary can be described by a linear combination of the Ishibashi states \cite{Cardy2004},
\begin{equation}
\ket{B} = \sum_i b_i \kett{i}.
\end{equation}
The explicit form of the Ishibashi state is
\begin{equation}
\ket{i}\rangle \equiv \sum_{N} \ket{i;N} \otimes   U \overline{\ket{i;N}},
\end{equation}
where $\ket{i;N}$ is a state in the Verma module $i$ labeled by $N$, and $U$ is an anti-unitary operator.
By unfolding the Ishibashi states (which we will denote by $||i|| \equiv \sum_{N} \ket{i;N} \otimes   U \overline{\bra{i;N}}$), we obtain an interface-like CFT where the interface-like operator $I(B) =  \sum_i b_i ||i||$ inserted along the line $z=0$ (see the right of FIG. \ref{fig:unfolding}).
In this picture,
the LRMI is just the mutual information between $A_1$ and $A_2$ ($A_i$ is defined in the right of FIG. \ref{fig:unfolding} ) in this interface-like CFT.

For the later use, we first show the calculation of the LREE for the whole system in our language.
The LREE for a system in a strip with size $L$ can be evaluated by the boundary primary correlator with two chiral twist operators,
\begin{equation}\label{eq:chiral}
S^{(l/r)}
=
\lim_{n \to 1}
\fr{1}{1-n}
\log
\fr{
\braket{\sigma^{b}_n(0) \bar{\sigma}^{b}_n(L)}_{strip}
}
{\braket{\mathbb{I}}_{strip}^n
,}
\end{equation}
where we denote the boundary primary by the superscript $b$.
The coefficient of the two-point function can be evaliated by the conformal map to a cylinder \cite{Das2015} as
\begin{equation}\label{eq:alpha}
\alpha_{\sigma_n^b} \equiv \fr{ \sum_i \abs{b_i}^{2n} S_{i0} }{ \pa{ \sum_i \abs{b_i}^{2} S_{i0}  }^n },
\end{equation}
where $S_{ij}$ is the modular S matrix and $h_n$ is the conformal dimension of the twist operator $h_n \equiv\fr{c}{24}\pa{n-\fr{1}{n}}$.
While one sometimes includes the cutoff parameter $\epsilon^{2h_n}$ into the coefficient,
we split this contribution from the coefficient in this article.
For example, in a diagonal RCFT, we obtain
\begin{equation}\label{eq:RCFT}
S^{(l/r)}
=
\fr{c}{6}\log\fr{L}{\epsilon} - \pa{\sum_i S_{ai}^2 \log \fr{S_{ai}^2 }{S_{0i}} },
\end{equation}
where we label the Cardy state by $a$. The constant term is called the topological entanglement entropy.
For special boundary states, the same constant term can also be found in (2+1)-$d$ TQFTs as the topological entanglement entropy \cite{Das2015}.

Let us move on to the LREE for a subsystem $A=[a,b]$, that is, the LRMI.
Except for some special models, the calculation of the LRMI is difficult.
To show a concrete calculation of the LRMI, we focus on the holographic CFT.
The entropy $S(A)$ can be evaluated by a correlation function of four chiral twist operators with the interface-like operator $I(B)$,
\begin{equation}\label{eq:unfoldMI}
\braket{ \bar{\sigma}_n( \bar{z}_1)  \sigma_n(\bar{z}_2)  I(B)  \sigma_n(z_1) \bar{\sigma}_n (z_2) }.
\end{equation}
This correlation function can be expanded as
\begin{equation}
g^n \sum_P C_{\sigma_n P}^2  \ca{F}^{Vir}_P(1-z),
\end{equation}
where the sum runs over boundary primaries and $\ca{F}^{Vir}_P(z)$ is the Virasoro block with the cross ratio $z\equiv \fr{z_{12}z_{34}}{z_{13}z_{24}}$.
The $g$-function represents a disk partition function \footnote{This $g$-function is not defined in the orbifold CFT $\ca{C}^{\otimes n}/\mathbb{Z}_n$ but in the seed CFT $\ca{C}$.}.
In this holographic CFT, the sum can be approximated by just the vacuum block if $z$ is enough large.
The bulk-boundary OPE coefficient $C_{\sigma_n \mathbb{I}}^2 $ can be evaluated by the unwrapping procedure \cite{Lunin2001}.
Note that the unwrapping procedure dose not change the profile of the boundary \cite{Akal:2021foz}.
The entropy $S^{(L)}_A$ is completely fixed by the conformal symmetry.
As a result, we obtain
\begin{equation}
\begin{aligned}
I^{(l/r)}(A) = \max \pa{ \fr{c}{6}\log \fr{(b-a)^2}{4ab}  - 2 \log g, 0 }.
\end{aligned}
\end{equation}
The trivial case, $I^{(l/r)}(A) =0$, is given by the vacuum block approximation of the dual-channel.

\section{Left-Right Reflected Entropy}
In a similar way to the LRMI, one can introduce another related quantity, {\it left-right reflected entropy (LRRE)},
which is a generalization of reflected entropy introduced in \cite{Dutta2021} (a similar notion was introduced in Chern-Simons theories \cite{Berthiere2021}).
To define the LRRE, we consider a canonical purification of a state $\rho_A$ in a doubled Hilbert space $\ca{H}_A \otimes \ca{H}^*_A$.
The LRRE is defined by the reduced density matrix obtained from the purified state $\ket{\sqrt{\rho_A}}$ by tracing over the right moving sector,
\begin{equation}
S^{(l/r)}_R(A) \equiv -\tr \rho^{(L)}_{AA^*} \log \rho^{(L)}_{AA^*},
\end{equation}
where $\rho^{(L)}_{AA^*}$ is the reduced density matrix of $\rho_{AA^*}=\ket{\s{\rho_{A}}} \bra{\s{\rho_{A}}}$  after tracing over the right moving sector.
The physical interpretation of the LRRE is similar to the LRMI.
For example, if one evaluates the LRRE in an integrable system, the LRRE behaves in the same way as the LRMI shown in FIG.\ref{fig:local}.
Nevertheless, if one focuses on a non-equilibrium process in a chaotic system where the quasi-particle picture breaks down, the LRRE shows a behavior different from the LRMI (see \cite{Kusuki2021,Kusuki2020,KudlerFlam2021,KudlerFlam2020}).

\begin{figure}[t]
 \begin{center}
  \includegraphics[width=8.0cm,clip]{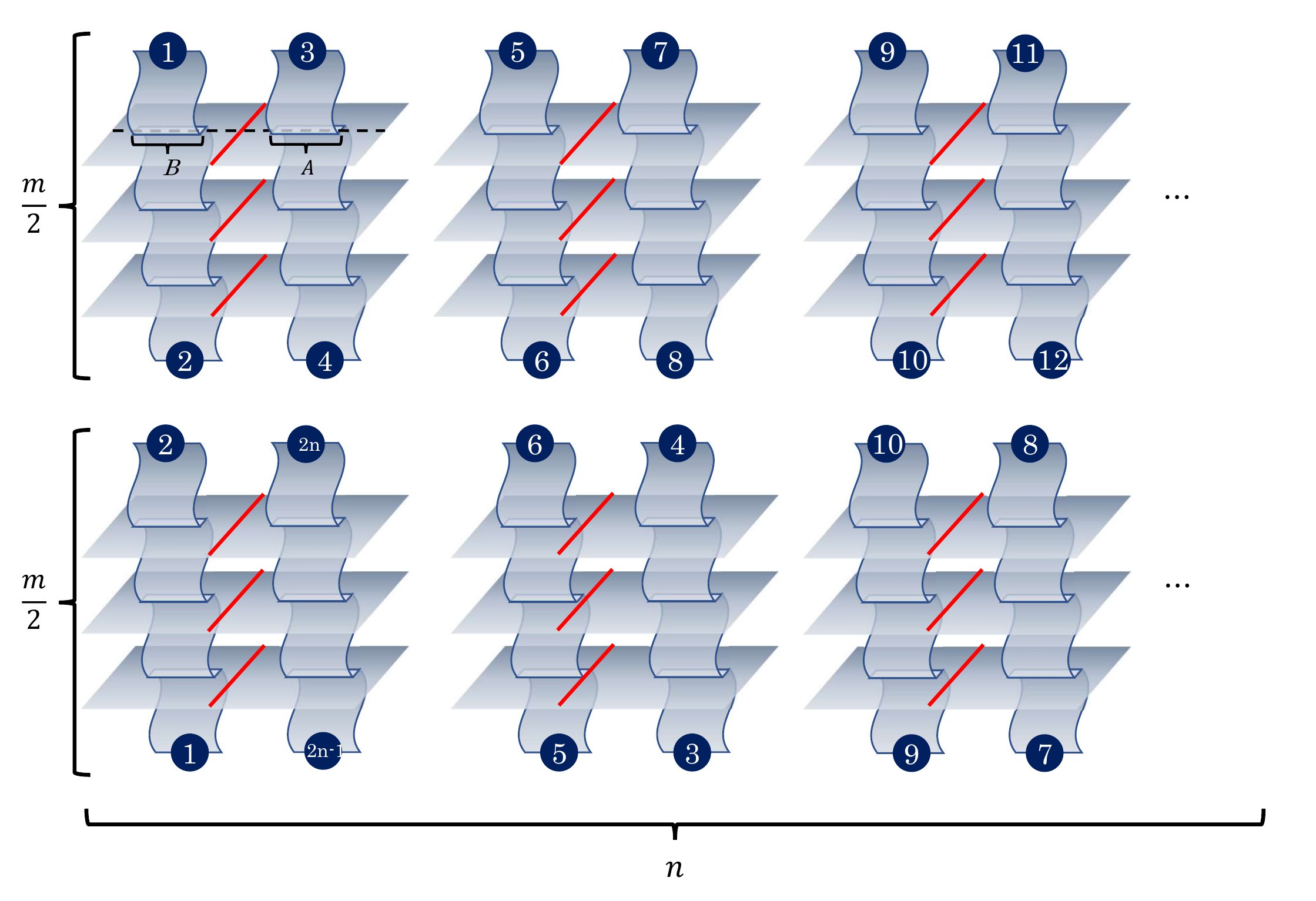}
 \end{center}
 \caption{The replica manifold that calculates the reflected entropy.  Edges labeled with the same number get glued together.
 The red lines describe the interface-like operators.
 }
 \label{fig:def}
\end{figure}

Let us focus on CFTs with time-like boundary at $z=0$.
To calculate the LRRE,
we can employ the replica trick in the path integral formalism as in \cite{Calabrese2004}.
In the same way as the LRMI,
we define the corresponding replica manifold in BCFTs by unfolding (see FIG.\ref{fig:unfolding}).
This replica manifold is shown in FIG. \ref{fig:def},
The reflected entropy can be evaluated by the partition function on this replica manifold  $Z_{n,m}$ as
\begin{equation}
S^{(l/r)}_R(A)=
\lim_{n,m \to 1} \fr{1}{1-n} \log
	\fr{Z_{n,m}}
	{\pa{ Z_{1,m}}^n },
\end{equation}
where the analytic continuation $m \to 1$ is taken for even integer $m$.
In a similar way to \cite{Dutta2021}, this replica partition function can be re-expressed as a correlation function with four chiral twist operators,
\begin{align}\label{eq:Znm}
&Z_{n,m}=\Big\langle\sigma_{g_A}(-b)\sigma_{g_A^{-1}}(-a)     I(B)  \sigma_{g_B}(a) \sigma_{g_B^{-1}}(b) \Big\rangle_{\text{CFT}^{\otimes mn}},
\end{align}
where we take the interval  $B=[a, b]$ and its mirror $A=[-b, -a]$.
For the same reason as (\ref{eq:unfoldMI}), the correlation function includes the interface-like operator.
To avoid unnecessary technicalities, we do not show the precise definition of the twist operators $\sigma_{g_A}$ and $\sigma_{g_B}$ (see \cite{Dutta2021}) because in this article, we only use the conformal dimension of the twist operators,
\begin{equation}
\begin{aligned}
&h_{\sigma_{g_A}}=h_{\sigma_{g_A^{-1}}}=h_{\sigma_{g_B}}=h_{\sigma_{g_B^{-1}}}=\fr{cn}{24} \pa{m-\fr{1}{m}}  (= n h_m)   ,\\
&h_{\sigma_{g_A^{-1} g_B}} = h_{\sigma_{g_B^{-1} g_A}}= \fr{c}{12}\pa{n-\fr{1}{n}} (= 2h_n),
\end{aligned}
\end{equation}
where $ h_{\sigma_{g_A^{-1} g_B}} $ ($ h_{\sigma_{g_B^{-1} g_A}} $) appears as the conformal dimension of the lowest primary operator in the OPE between $\sigma_{g_A^{-1}} $ and $\sigma_{g_B}$ ( $\sigma_{g_B^{-1}}$ and $ \sigma_{g_A}$ ).

\begin{figure}[t]
 \begin{center}
  \includegraphics[width=8.0cm,clip]{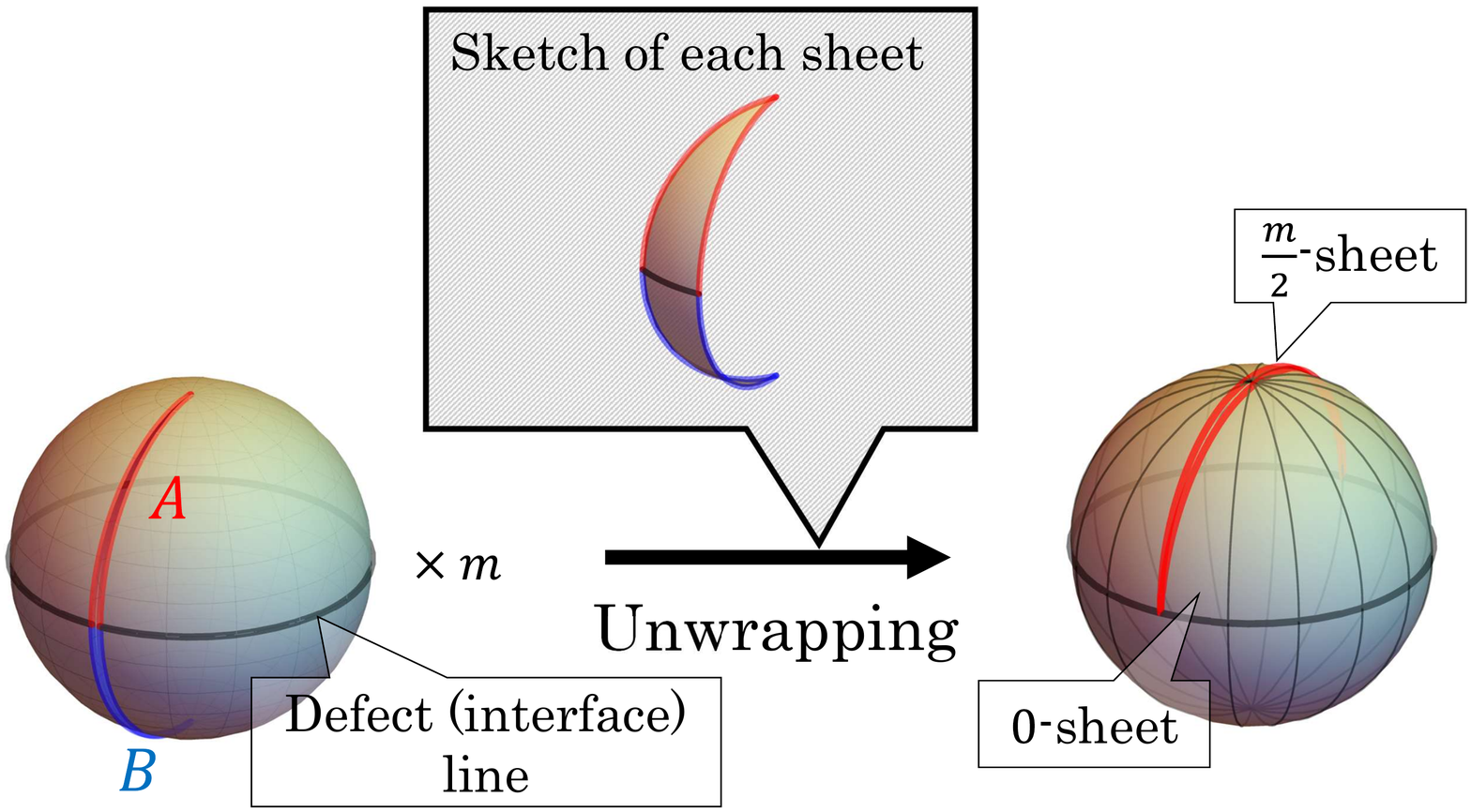}
 \end{center}
 \caption{Sketch of how to evaluate the OPE coefficient.
 We can unwrap the $m$-fold branch cut by the conformal transformation $z\to z^{\fr{1}{m}}$.
 This unwrapping leads to a sphere with one interface line (black line) and the $n$-fold branch cut (red line), whose edge is attached to the interface.
 This is completely the same as the setup studied in \cite{Sakai2008,Brehm2015}.
 Using the technique in  \cite{Sakai2008,Brehm2015}, one can evaluate  this partition function.
 }
 \label{fig:unwrapping}
\end{figure}

\begin{figure}[t]
 \begin{center}
  \includegraphics[width=5.0cm,clip]{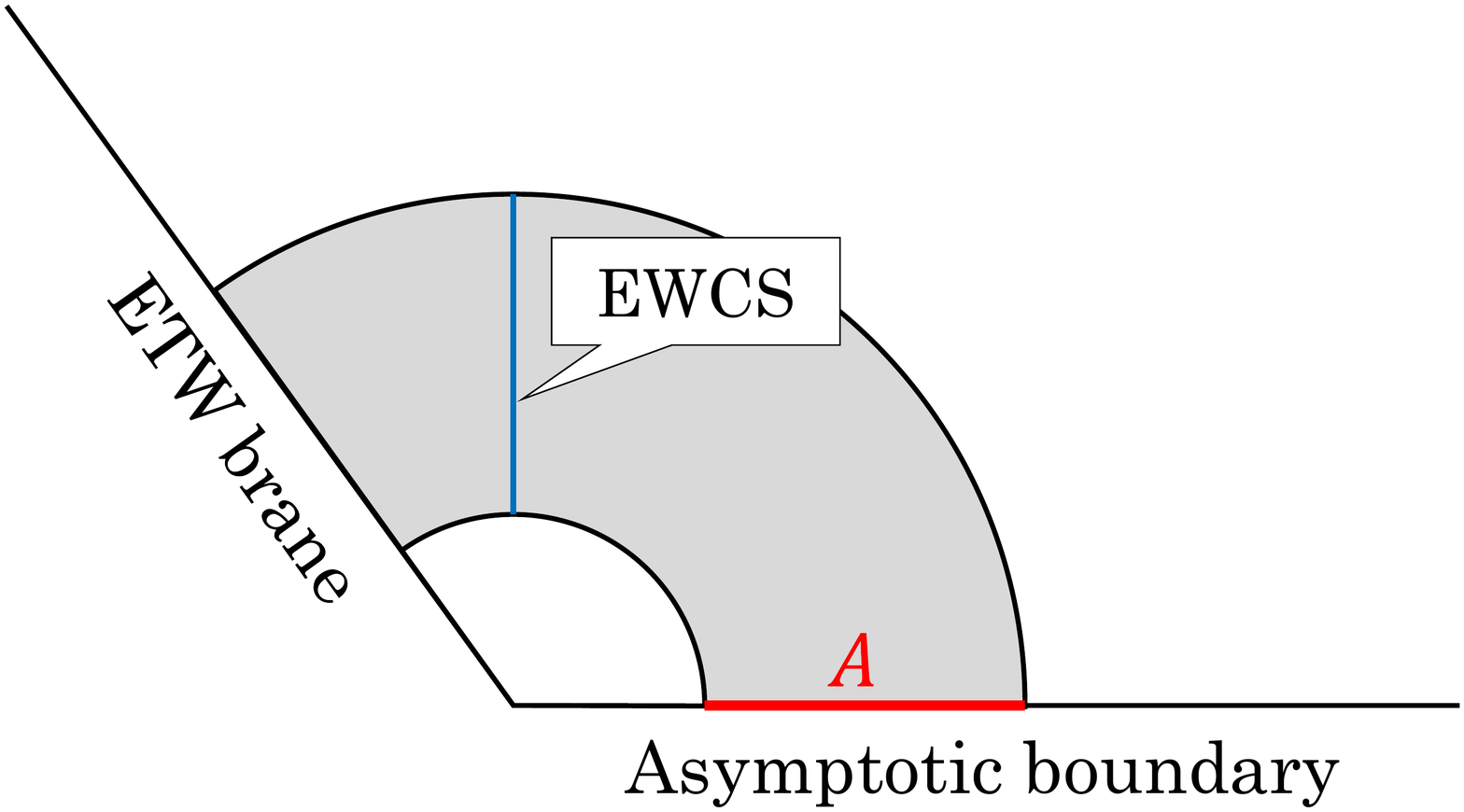}
 \end{center}
 \caption{The holographic dual of the LRRE in BCFTs. This is given by the entanglement wedge cross section (EWCS), which is defined in  \cite{Takayanagi2018,Nguyen2018}.}
 \label{fig:EWCS}
\end{figure}

The reflected entropy has a nice bulk interpretation.
One can show that the reflected entropy is dual to twice the area of the entanglement wedge cross section,
$S_R=2E_W$  \cite{Dutta2021}.
In a similar way, one can find the holographic dual of the LRRE.
In the holographic CFT,
the correlation function (\ref{eq:Znm}) in the limits $n\to1$ and $m \to 1$ can be approximated by a single Virasoro block $\ca{F}^{Vir}(2h_n|1-z)$ if $z$ is enough large and then we have
\begin{equation}\label{eq:Z_nm}
\begin{aligned}
	\fr{Z_{n,m}}
	{\pa{ Z_{1,m}}^n }
=  \alpha_{\sigma_n^b}^{-2}   C_{\sigma_{g_A^{-1}} \sigma_{g_B} \sigma_{g_A g_B^{-1}}^b} ^2  \pa{\fr{1+\sqrt{z}}{2\sqrt{1-z}}}^{-4h_n}.
\end{aligned}
\end{equation}
The factor $ \alpha_{\sigma_n^b}^{-2}$ comes from the normalization of the twist operator (\ref{eq:alpha}).
The OPE coefficient $ C_{\sigma_{g_A^{-1}} \sigma_{g_B} \sigma_{g_A g_B^{-1}}^b} $ can be evaluated by the unwrapping trick \cite{Lunin2001} (see also \cite{Dutta2021}).
Since the unwrapping procedure does not affect the interface,
we have
\begin{equation}\label{eq:OPE}
C_{\sigma_{g_A^{-1}} \sigma_{g_B} \sigma_{g_A g_B^{-1}}^b} =
\pa{2m}^{-2h_n} \alpha_{\sigma_n^b},
\end{equation}
where $\alpha_{\sigma_n^b}$ is defined in (\ref{eq:alpha}). The reason why we obtain the coefficient  $\alpha_{\sigma_n^b}$ is explained in FIG.\ref{fig:unwrapping}.
Thus, the reflected entropy is given by
\begin{equation}\label{eq:Slr}
S^{(l/r)}_R(A)
=
\fr{c}{3} \log \pa{\fr{b}{a}} + const.,
\end{equation}
which completely matches twice the area of the entanglement wedge cross section defined in FIG.\ref{fig:EWCS} up to constant.

One can also evaluate the LRRE in the adjacent limit.
Let us first consider the limit $1-z = \fr{4ab}{(a+b)^2} \ll \epsilon \ll 1$.
In this case, the replica partition function can be approximated by 
\begin{equation}
\begin{aligned}
	\fr{Z_{n,m}}
	{\pa{ Z_{1,m}}^n }
=   C_{\sigma_{g_A^{-1}} \sigma_{g_B} \sigma_{g_A g_B^{-1}}^b}   \pa{ \fr{ (a+b)^2}{ab}  }^{-2h_n}.
\end{aligned}
\end{equation}
As a result, the LRRE for the Cardy state $\ket{a}$ in a diagonal RCFT is given by
\begin{equation}
S^{(l/r)}_R(A)
\to
\fr{c}{6}\log{\fr{ (a+b)^2}{ab} }  - 2\pa{\sum_i S_{ai}^2 \log \fr{S_{ai}^2 }{S_0i} }.
\end{equation}
The second term comes from the coefficient $\alpha_{\sigma_n^b}$ in (\ref{eq:OPE}).
On the other hand, if we take the limit $\epsilon \ll 1-z = \fr{4ab}{(a+b)^2}  \ll 1 $,
we obtain from (\ref{eq:Z_nm})
\begin{equation}
S^{(l/r)}_R(A)
\to
\fr{c}{6}\log{\fr{ (a+b)^2}{ab} }.
\end{equation}
In the limit of the adjacent intervals, one can find that the Markov gap \cite{Hayden2021} has the following form,
\begin{equation}\label{eq:markov}
S^{(l/r)}_R(A) - I^{(l/r)}_R(A)=\fr{c}{3}\log2+\cdots.
\end{equation}
This universal term can also be found in a special tripartition setup \cite{Zou2021}.
The additional terms $\cdots$ depend on the details of the boundary.

Note that in the holographic CFT, the LRRE for a finite subsystem satisfies the inequality (an analog of \cite{Hayden2021} in CFTs without boundary),
\begin{equation}
S^{(l/r)}_R(A) - I^{(l/r)}_R(A) \geq \fr{c}{6}\log2 \times (\text{\# of cross section bounraries}).
\end{equation}
It is claimed that the reflected entropy is more sensitive to multipartite entanglement \cite{Akers2020, Zou2021}.
The $O(c)$ difference between the reflected entropy and the mutual information implies that there must be a large amount of tripartite entanglement
in our tripartition setup associated with the division of the left/right moving sectors.

\section{Reflected Entropy in Interface CFT}

\begin{figure}[t]
 \begin{center}
  \includegraphics[width=5.0cm,clip]{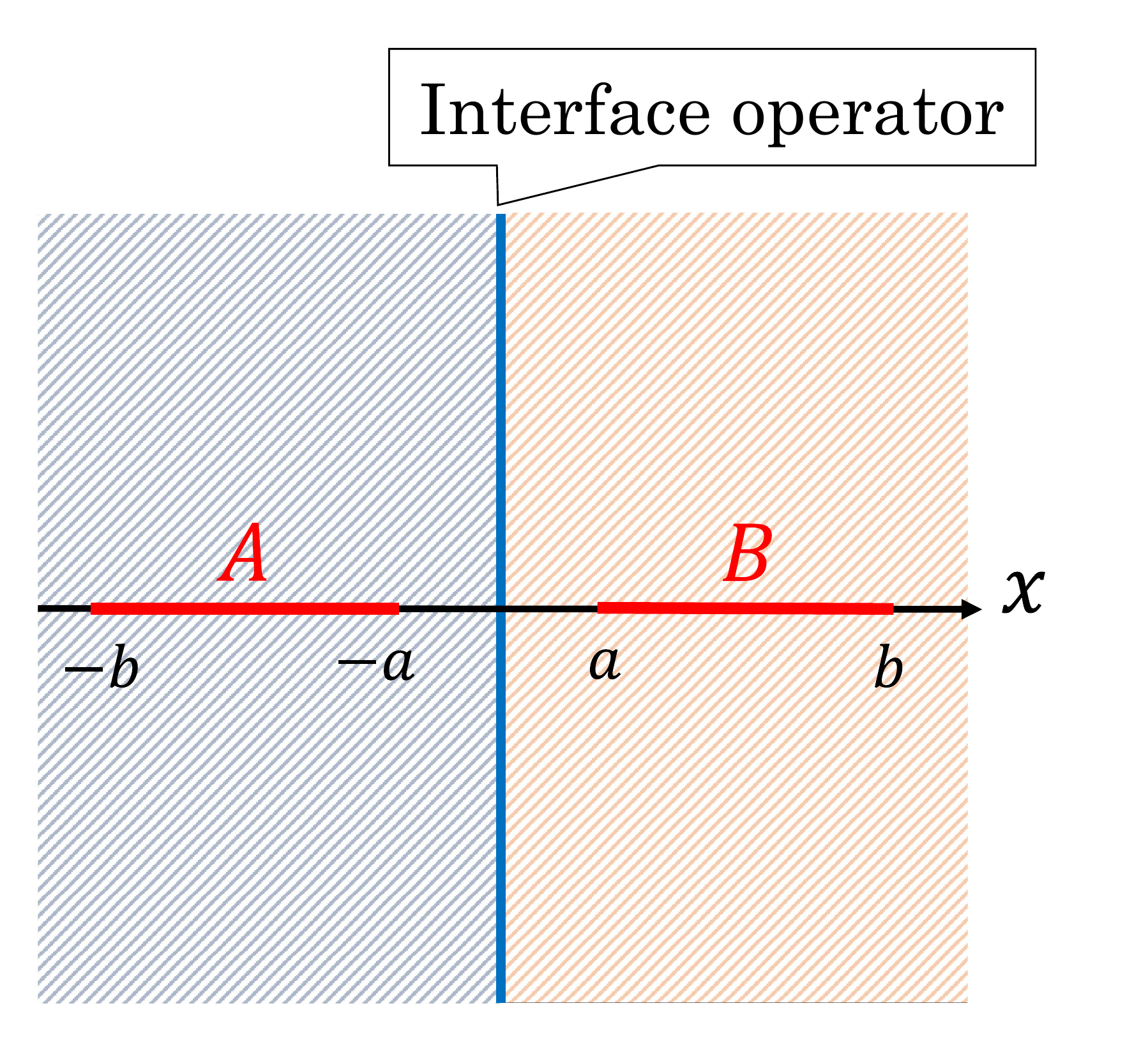}
 \end{center}
 \caption{The setup that we consider. The interface is symmetrically localed on the two intervals. One can also consider an asymmetric setup for topological interfaces.}
 \label{fig:icft}
\end{figure}

In a similar way to the LRRE, we can define the reflected entropy in an interface CFT, i.e., $\text{CFT}_1 \otimes \overline{\text{CFT}}_2  $ with central charge $c_1$ and $c_2$
(see our setup in FIG.\ref{fig:icft}),
\begin{equation}
S_R(A,B)=
\lim_{n,m \to 1} \fr{1}{1-n} \log
	\fr{Z_{n,m}}
	{\pa{ Z_{1,m}}^n },
\end{equation}
where the partition function is expressed in terms of (not chiral) twist operators as
\begin{align}
&Z_{n,m}=\Big\langle\sigma_{g_A}(u_1)\sigma_{g_A^{-1}}(v_1)     I(A)  \sigma_{g_B}(u_2) \sigma_{g_B^{-1}}(v_2) \Big\rangle_{\text{CFT}^{\otimes mn}},
\end{align}
where we take the intervals $A=[u_1,v_1]$ and $B=[u_2,v_2]$ with $u_1<v_1<u_2<v_2$.
Since the calculation of the reflected entropy is still difficult in general,
we focus on the case where the correlation function is approximated by the single block, as in the holographic CFT or in the limit of the adjacent intervals.
In a similar way to the BCFT case,
one may approximate the correlation function in the holographic CFT as 
\begin{equation}\label{eq:icftSR}
\begin{aligned}
	\fr{Z_{n,m}}
	{\pa{ Z_{1,m}}^n }
&= C_{\sigma_{g_A^{-1}} \sigma_{g_B} \sigma_{g_A g_B^{-1}}^b} ^2 \abs{ \ca{F}^{Vir}(2h'_n|1-z) }^2,
\end{aligned}
\end{equation}
where $h'_n \equiv \fr{c_{\text{eff}}}{24}\pa{n-\fr{1}{n}} $
\footnote{This conformal dimension can be obtained by evaluating the vacuum energy of the closed string sector of the cylinder partition function associated with the replica manifold shown in the right of FIG.\ref{fig:unwrapping}. The same technique can be found in \cite{Sakai2008,Brehm2015}.}
and
\begin{equation}
C_{\sigma_{g_A^{-1}} \sigma_{g_B} \sigma_{g_A g_B^{-1}}^b} =
\pa{2m}^{-4h_n} \alpha_{\sigma_n^b}^2,
\end{equation}
where the twist operator is not a chiral twist operator unlike that in (\ref{eq:chiral}).
This is true for topological defects but in general, could not be true.
This is because one cannot organize the descendant contributions in the same way as CFT without defects.
One thing we can say for sure is that the leading contribution in the limit of the adjacent intervals $z\to 1$ is given by
\begin{equation}
\begin{aligned}
	\fr{Z_{n,m}}
	{\pa{ Z_{1,m}}^n }
&=  \alpha_{\sigma_n^b}^{-2}  C_{\sigma_{g_A^{-1}} \sigma_{g_B} \sigma_{g_A g_B^{-1}}^b} ^2 \abs{ (1-z)^{2h'_n}  }^2.
\end{aligned}
\end{equation}
The effective central charge $c_{\text{eff}} \in [0,\min(c_1,c_2)]$ depends on a profile of the interface (see \cite{Sakai2008}).
Consequently, the reflected entropy in the limit of the adjacent intervals $1-z = \fr{4ab}{(a+b)^2} \ll 1 $ is given by
\begin{equation}\label{eq:SR2}
S_R(A,B)=
\fr{c_{\text{eff}}}{3} \log{\fr{ (a+b)^2}{ab} }   + const..
\end{equation}
The OPE coefficient can be obtained by the unwrapping procedure (see FIG.\ref{fig:unwrapping}).
For the same reason as the entanglement entropy,
it is difficult to evaluate the OPE coefficient (more precisely, an analog of $\alpha_{\sigma_n^b}$) in generic interface CFTs.
Nevertheless, in a specific case, topological interface, we can give the explicit form of the constant part in a similar way as the LRRE in BCFTs.
(See \cite{Brehm2016} for the derivation of $\alpha_{\sigma_n^b}$ and its interpretation.)
For topological interfaces,
the Markov gap is given by
\begin{equation}
S_R(A,B)-I(A,B)  =  \fr{2c}{3} \log 2 + \cdots.
\end{equation}
The additional terms $\cdots$ depend on the details of the interface.
This universal term is completely consistent with  \cite{Zou2021}.
In general interfaces, the Markov has a complicated form.
We provide further analysis in \cite{KudlerFlam}.

Let us focus on the holographic interface.
One natural realization of the holographic dual of the interface CFT is given by the so-called Janus solution \cite{Bachas2002,Bak2003,Bak2007}.
The entanglement entropy in this holographic interface is evaluated in various setups \cite{Azeyanagi2008, Gutperle2016, Karch2021}.
In a similar way to \cite{Azeyanagi2008, Gutperle2016, Karch2021}, the entanglement wedge cross section can be evaluated.
In the case considered in \cite{ Karch2021} (also in \cite{Azeyanagi2008, Gutperle2016}),
the reflected entropy is given by
\begin{equation}\label{eq:holo1}
S_R(A,B)
=
\fr{2c_{\text{eff}}}{3} \log \pa{\fr{b}{a}} + const..
\end{equation}
The effective central charge is the same as that found in \cite{Gutperle2016}.

There is another realization by simply connecting two geometries by a thin brane \cite{Azeyanagi2008,Karch2021,Erdmenger2015,Simidzija2020,Bachas2020, Bachas2021,Bachas2021a, Anous2022}, which is a generalization of the bottom-up AdS/BCFT \cite{Takayanagi2011,Fujita2011}.
Although there is limited knowledge about the CFT dual of this model, this model is interesting for two reasons.
Unlike the top-down model, the thin-brane model is a minimal gravity dual of ICFT.
This is an analog of pure gravity.
Another reason comes from the island model \cite{Penington2020,Almheiri2019,Almheiri2020}.
One can think of the thin-brane as a gravity theory coupled to a non-gravitational bath CFT, called the braneworld holography \cite{Karch2001}.
Through this holography, one can investigate the island model by the ICFT dual to the thin-brane model.
In the thin-brane model, the minimal cross section of the entanglement wedge is given by
\begin{equation}\label{eq:holo2}
S_R(A,B)
=
\fr{2 \min(c_1,c_2)}{3} \log \pa{\fr{b}{a}}  + const..
\end{equation}
This form of the reflected entropy is non-trivial as we mentioned below (\ref{eq:icftSR}).
This may strongly constrain the profile of the holographic interface.

\section{Discussions}
In this article, we develop the idea of reflected entropy in BCFT/ICFT.
We propose some remaining questions and interesting future works.
It would be interesting to investigate the reflected entropy in various setups with boundaries or interfaces,
which would tell us about the multipartite entanglement between bulk and boundary/interface.
In the holographic CFT, one can explicitly evaluate the reflected entropy in BCFT/ICFT,
which has the potential to identify the profile of the holographic interface in the CFT language.
An interesting future work is to apply our analysis to the island model, which is essentially a special class of BCFT/ICFT.
From such an analysis, one may be able to understand the multipartite entanglement structure of the island model.
It is known that the LREE in (1+1)-$d$ CFT has a nice interpretation in (2+1)-$d$ TQFT \cite{Das2015, Berthiere2021, Nishioka2021}.
It would be interesting to find the TQFT picture of the LRRE.

\section*{Acknowledgments}
We thank Jonah Kudler-Fram, Yuhan Liu, Shinsei Ryu, Ramanjit Sohal, Kotaro Tamaoka, and Zixia Wei for fruitful discussions and comments.
YK is supported by Burke Fellowship (Brinson Postdoctoral Fellowship).

\bibliography{main}

\if(
\pagebreak
\widetext
\begin{center}
\textbf{\large Supplemental Materials}
\end{center}
\setcounter{equation}{0}
\setcounter{figure}{0}
\setcounter{table}{0}
\setcounter{page}{1}
\makeatletter
\renewcommand{\theequation}{S\arabic{equation}}
\renewcommand{\thefigure}{S\arabic{figure}}
\renewcommand{\bibnumfmt}[1]{[S#1]}
\renewcommand{\citenumfont}[1]{S#1}

\section{Left-Right Entanglement Entropy for the whole system}

)\fi

\end{document}